\documentclass[%
 preprint,
 amsmath,amssymb,
]{revtex4-1}

\setcitestyle{authoryear,open={(},close={)}}

\usepackage{graphicx}
\usepackage{dcolumn}
\usepackage{bm}
\usepackage{hyperref}
\usepackage{amsmath}
\usepackage{multirow}
\usepackage{url}

\usepackage[utf8]{inputenc}
\usepackage{booktabs}
\usepackage{setspace}
\usepackage{subcaption,booktabs}

\usepackage{color}
\definecolor{redcolor}{rgb}{1.0,0.,0.}

\begin{document}

\preprint{}

\title{On predicting research grants productivity}
\author{Jorge A. V. Tohalino}%
\author{Diego R. Amancio}%
\email{diego@icmc.usp.br}
\affiliation{%
Institute of Mathematics and Computer Science, Department of Computer Science, University of S\~{a}o Paulo,
S\~{a}o Carlos, SP,  Brazil
}%

\date{\today}
\begin{abstract}
Understanding the reasons associated with successful proposals are of paramount importance to improve evaluation processes. In this context, we analyzed whether bibliometric features are able to predict the success of research grants. We extracted features aiming at characterizing  the academic history of Brazilian researchers, including research topics, affiliations, number of publications and visibility.
The extracted features were then used to predict grants productivity via machine learning in three major research areas, namely Medicine, Dentistry and Veterinary Medicine. We found that research subject and publication history play a role in predicting productivity. In addition, institution-based features turned out to be relevant when combined with other features. While the best results outperformed text-based attributes, the evaluated features were not highly discriminative. Our findings indicate that predicting grants success, at least with the considered set of bibliometric features, is not a trivial task.
\end{abstract}


\maketitle

\section{Introduction}\label{section:intro}

In recent years, \emph{Science of Science} emerged as an important application of big data analysis~\citep{Fortunatoeaao0185}. Owing to the availability of large data sets derived from the scientific literature, several studies have been conducted to shed light on how science is organized and evolves as a complex system~\citep{zeng2017science}.
Examples of approached topics include science evolution~\citep{silva2016using}, collaboration/citation patterns and measures to evaluate science~\citep{bar2008h}. More recently, studies in \emph{Science of Science} have also focused on predictive tasks, which has become very important in different scenarios~\citep{acuna2012predicting}. For example, automatic approaches have been used to predict when a new topic will emerge~\citep{salatino2017topics}. In a similar fashion, neural networks representations have been able to predict outcomes of scientific research~\citep{bagrow2018neural}. Mobility trajectories of researchers have also been studied using computational methods~\citep{he2019measuring}. Equally important are those studies predicting scientific success, including the prediction of papers and scholars' impact~\citep{wang2019early}.

More recently, several studies have focused on analyzing the factors underlying grants success~\citep{boyack2018toward,tohalino2020analyzing,letchford2016advantage,paiva2012articles}. Understanding the factors that may lead to successful grants are ultimately important to determine which proposals are the most promising and relevant to be funded. While machine-based techniques are not meant to replace an expert, thorough analysis of proposals, they may assist the analysis of a large number of documents and other metadata extracted from research proposals. Potential advantages associated with the use of machine learning methods to assist the analysis of proposals include an analysis less prone to personal bias, and a much faster review compared to traditional human evaluations. In addition, automatic analyses could also be used to understand the factors correlated to successful grants.

Text- and reference-based features have been used to predict the success of research proposals~\citep{boyack2018toward,tohalino2020analyzing,letchford2016advantage,paiva2012articles}. \cite{boyack2018toward}  found that  proposals success depends on the topic being approached. More specifically, \cite{boyack2018toward} found that subjects that have already been studied by the researcher are more likely to yield successful grants. The topic similarity in this case was computed by comparing proposal references and the respective applicant publications. A text analysis was conducted by~\cite{markowitz2019words}. The authors studied if text complexity measurements extracted from NSF projects correlate with the amount of funding received by the researchers. They found that larger abstracts comprising a low number of common words are among the main patterns associated with larger funding values. In a similar study, \cite{tohalino2020analyzing} found that topical and complexity textual features play a role in grants predicting grants productivity, but the prediction values were not very high.

Different from other approaches, here we use machine-learning methods applied to features extracted from researchers, institutions and publications to analyze whether those features can be used to predict the productivity of grants. We used several features such as total number of publications, citations, relevance of PI's institution and diversity of the approached subjects. Given the unbalanced nature of the dataset, we considered as criteria for productivity the publication of at least one scientific paper~\citep{tohalino2020analyzing}. Using a dataset of research grants from the \emph{São Paulo Research Foundation} (FAPESP-Brazil), our analysis was conducted in three distinct research areas, namely Medicine, Dentistry and Veterinary Medicine.

Several interesting results have been found from our analysis. The analysis based on classifiers with a single feature showed that there is a relationship between the studied features and future productivity of grants. In this single-feature analysis, features based on research subjects and on publication/citation counts were the most effective to predict grants productivity. The analysis combining different features in the same classifier also showed an improvement in performance. The highest accuracy rates were found for the Veterinary Medicine area. In this case, we could discriminate productive grants with an accuracy higher than $67\%$. While the results are significant, the typical prediction accuracy were not very high. They were typically higher, though, than approaches based on textual features alone~\citep{tohalino2020analyzing}. Our analysis also revealed that both \emph{Support Vector Machines} and \emph{Multilayer Perceptron} were the classifiers yielding the highest accuracy rates.
Despite being a challenging task, we believe that the studied features could be combined with additional information to allow a better understanding of the factors correlated with grants success.

This manuscript is organized as follows.
In Section~\ref{section:methodology}, we describe the methodology, including the description of features and the machine learning framework. We discuss the obtained results in Section~\ref{section:results}. Finally, in Section~\ref{sec:conc}, we present the conclusions and perspectives for future works.

\section{Material and methods}\label{section:methodology}

In order to classify research grants according to their productivity, the following steps were taken:

\begin{enumerate}

    \item \emph{Dataset collection}: the dataset we used comprises research projects supported by \emph{S\~ao Paulo Research Foundation} (FAPESP-Brazil). The dataset is available from the \textit{Biblioteca Virtual} website (see Section~\ref{methodology:dataset}). In addition to the information regarding research projects (number of papers derived from the grant, title, abstract etc), the dataset also provides information to characterize PIs (e.g their publication history) and institutions (e.g. universities and research institutes).

    \item \emph{Feature extraction}: this step is responsible for extracting features from researchers that are used to predict grants productivity. Our hypothesis is that the success of a grant could be  dependent on PIs features, such as previous success in other grants and publication/citation history. Several features were extracted to characterize authors. Examples of extracted features are: number of funded projects, number of publications and citations yielded by the researcher's grants, affiliation and diversity of subfields studied by the researcher. Section~\ref{methodology:features} describes the features we used to perform the classification.

    \item \emph{Classification}: the aim of this phase is automatically identify productive research proposal according to the established criteria for productivity. We considered a binary classification task. The features extracted from the previous step were used as input for traditional machine learning algorithms. We also performed several tests in order to find the best combination of features.
    In Section~\ref{methodology:classification}, we describe the classification step. This phase includes the training and evaluation phase.

\end{enumerate}

\begin{figure}[h]
    \centering
    \includegraphics[width=1.0\textwidth]{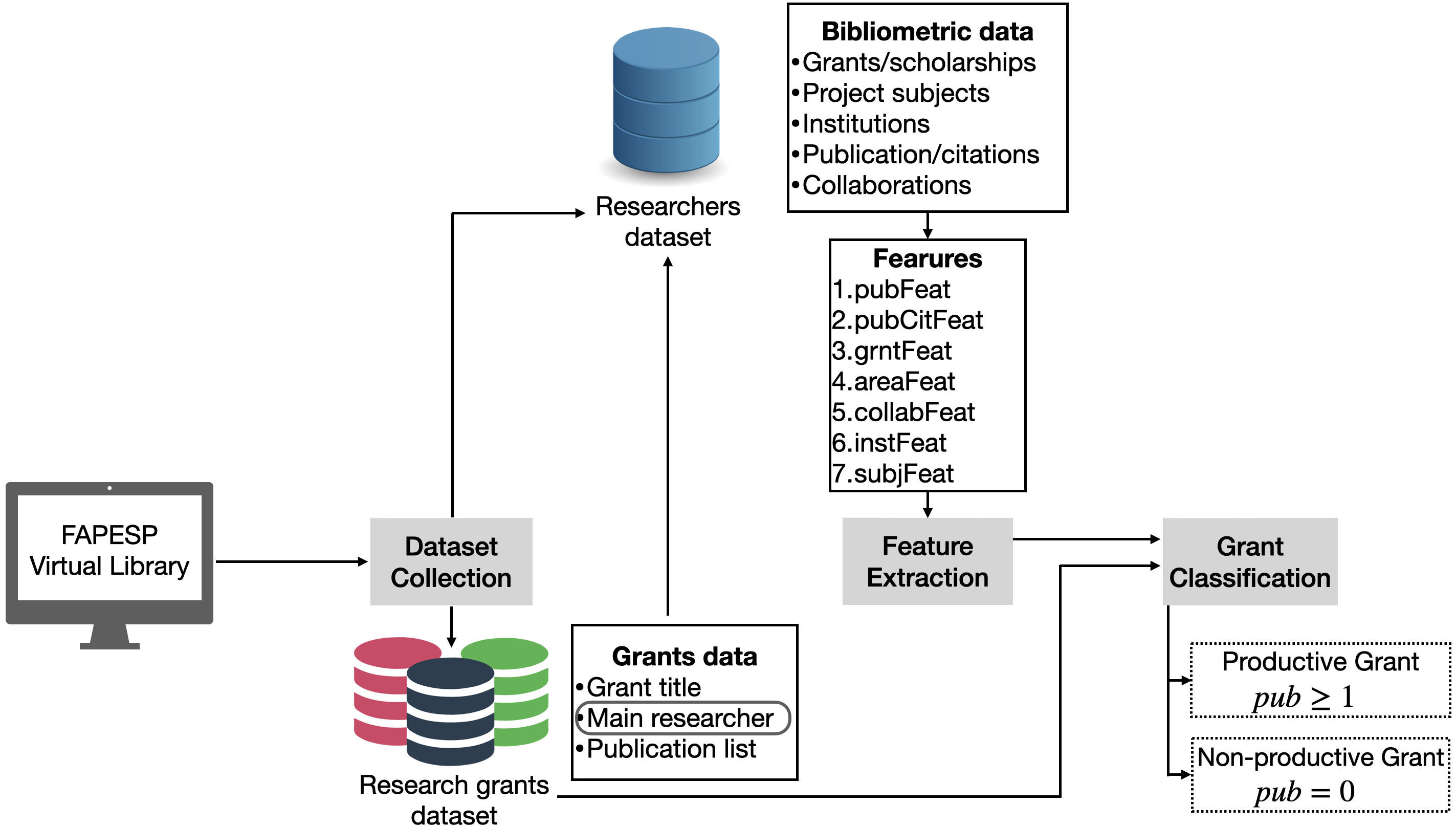}
    \caption{Architecture of the proposed methodology for research grant classification.
    }
   \label{fig:architecture}
\end{figure}

In Figure~\ref{fig:architecture}, the architecture for research grant classification is shown.  First, we collected relevant information from the FAPESP Dataset (FAPESP Virtual Library). This includes information from PIs that are related to their previous research experience and other features linked to their professional activity. All information from researchers are collected in the researcher dataset. Examples of features extracted are the number of publications obtained in previous grants (\emph{pubFeat}), number of citations received by these publications (\emph{pubCitFeat}) and other features that are detailed later on. These features are used to train supervised classifiers in a binary classification task to predict whether a grant will be productive. We use the number of publications resulting from the grant as the criterion to measure productivity.

\subsection{Dataset collection}\label{methodology:dataset}

The considered dataset comprises a subset of grants  offered by \emph{S\~ao Paulo Research Foundation} (FAPESP)~\citep{tohalino2020analyzing}. FAPESP is an important public research foundation in Brazil and is fully funded by the State of S\~ao Paulo~\footnote{https://fapesp.br/en/about}. The metadata regarding PIs and research grants were retrieved from \emph{Biblioteca Virtual} website~\footnote{https://bv.fapesp.br/pt/}. We focused our analysis on \emph{regular grants}, which are grants with an average duration of 18-24 months. All FAPESP regular grants are conducted under the supervision of a principal investigator, who must be associated with a university (or research institution) from S\~ao Paulo. We decided to study this type of grant because the \emph{Biblioteca Virtual} has a large number of regular grants (roughly 31,000 instances). We selected research grants starting before 2016 with duration between 23 and 24 months. Because recent grants were disregarded, all considered projects had at least 3 years to yield at least one publication after the grant is finished.

Each research grant has a list of associated publications. This information is to automatically extracted from the Web of Science (Clarivate Analytics) dataset by the \emph{Biblioteca Virtual} website. This automatic extraction is possible because any paper published in the context of a FAPESP research grant must acknowledge FAPESP in a specific format. Regular grants particularly
are acknowledged using the format ``yyyy/nnnnn-d'', where ``yyyy'' represents the year when the research project was submitted and ``nnnnn-d'' is the grant number.

Grants funded by FAPESP cover a wide variety of research areas, including, e.g.  Health, Biological and Earth Sciences. Because distinct areas have different publication patterns~\citep{rafols2010diversity}, we decided to compare grants \emph{within} the same field. Our analysis was conducted in the three largest fields: Medicine (MED), Dentistry (DENT) and Veterinary Medicine (VET).

\subsection{Feature extraction}\label{methodology:features}

For each research grant, we extracted several features related to the respective PI. The features are meant to characterize researchers' academic trajectory just before the grant started. The features used to characterize PIs were grouped into the following groups:

\begin{enumerate}

    \item \emph{Publication-based features} (\emph{pubFeat}): here we use features that are related to PIs publications. Our hypothesis here is that a good previous performance related to publications could be a good indicator of future performance~\citep{lu2019analyzing}. The following publication-based features were used for this analysis:
    \begin{enumerate}

        \item total number of publications;

        \item number of grants yielding at least one publication;

        \item maximum number of publications resulting from a single grant received by the PI;

        \item average productivity (in number of publications); and

        \item number of grants at least one publication divided by the total number of grants received by the PI.

    \end{enumerate}

    \item \emph{Features based on both publications and citations} (\emph{pubCitFeat}): While scientific publications denote researchers effort to provide new pieces of knowledge, citations can be considered as a metric of relevance and visibility~\citep{kong2020gene,siudem2020three}. Our hypothesis is that citations can be used as a proxy to PIs scientific influence~\citep{ioannidis2020updated}. Thus, we investigate whether  influential researchers are more likely to conduct productive research grants. The following PIs measurements were considered for this set of features:

    \begin{enumerate}

        \item total number of articles published;

        \item total number of citations accrued by the researcher;

        \item average number of publications per year; and

        \item average number of citations per year.

    \end{enumerate}

    \item \emph{Features based on the number of grants and scholarships received by the PI} (\emph{grntFeat}): our hypothesis is that more experienced researchers are more likely to have a productive grant~\citep{markowitz2019words,larrimore2011peer}. The degree of researcher's experience was measured in terms of the total
    number of grants and scholarships received by the researcher. In addition to regular grants, we also considered as features the number of undergraduate, master's, doctoral and post-doctoral degree scholarships supervised by the researcher.

    \item \emph{Features based on the diversity of research areas} (\emph{areaFeat}): our hypothesis here is that PIs might have experience on diverse research areas, and this could be an indication of future productivity. Some studies have shown that interdisciplinary journals and papers are more visible in the sense that they tend to  attract more citations than more specialized research~\citep{rinia2002impact}. In a similar fashion, our hypothesis is that interdisciplinary research could be more visible and this could facilitate the publication of papers since more journals and scholars could be interested in the interdisciplinary results being disseminated. The features used to quantify the degree of interdisciplinary encompasses three different granularity levels. We considered the number of areas in each of the first three levels. Examples of top-levels areas include
    Exact and Health Sciences. Examples of second-level hierarchy areas for Health Sciences include e.g. Nursing, Pharmacy, Medicine and Dentistry. Finally, examples of third level areas for Medicine include Medical Clinic, Maternal Health, Surgery and Psychiatry.

    \item \emph{Collaboration-based features} (\emph{collabFeat}): in this set of features we evaluate whether the number of different collaborators in the past might be correlated with grants success. A large number of collaborators could be a proxy to quantify researchers' experience (or even seniority) and thus collaboration-based features could indicate if an author is able to gather researchers with different backgrounds to conduct scientific research. Because more collaborations could be also correlated with more distinct contributions~\citep{correa2017patterns}, we could also expect that joint effort could be correlated with higher quality research~\citep{franceschet2010effect}, which in turn could positively contribute to the success of a research grant. The following features were used to quantify the diversity of PIs collaborations:

    \begin{enumerate}

        \item total number of local collaborators in research grants. Local researchers are all researchers affiliated to a Brazilian research (or higher-education) institutions;

        \item total number of abroad collaborators;

        \item total number of grants received by the PI with one or more associated researchers;

        \item total number of distinct co-authors in scientific publications; and

        \item average number of co-authors per article.

    \end{enumerate}

\item \emph{Institution-based features} (\emph{instFeat}):
institution-based features are used to probe whether PIs affiliation plays a role in predicting the success of research grants. The hypothesis is that grants conducted at larger (or more visible) institutions are more likely to yield a productive grant. More prestigious institutions could favour productivity given that more prestigious institutions could themselves host more productive researchers~\citep{bauder2020international}. In addition, more prestigious universities could also have more access to the material and resources to conduct high-quality research. The visibility and importance of institutes were measured in terms of the following features:

\begin{enumerate}

    \item total number of projects hosted by the PI's institution;

    \item total number of publications associated with the PI's institution; and

    \item total number of productive grants hosted by the PI's. We used here the criteria discussed in Section \ref{methodology:dataset} to classify a grant as productive.

\end{enumerate}

The set of features mentioned in (a)-(c) are henceforth referred to as $\emph{instFeat}_{A}$. We also considered an additional data representation, where we do not consider the features extracted from the institution, but the host institution becomes a feature. More specifically, a vector is used to represent if the PI belongs to a specific institution. The $i$-th element of the vector takes the value 1 if the PI is affiliated to the $i$-th institution. Otherwise, the value stored is zero. This representation is henceforth referred to as  $\emph{instFeat}_{B}$. We also used a different representation referred to as $\emph{instFeat}_{C}$. This representation uses a vector that is similar to the previous version, but instead of assigning the value $1$, we assigned the value of the success rate of the researcher's university or institution.

    \item \emph{Success of research subjects} (\emph{subjFeat}):
    each research project in the dataset comprises keywords (or keyphrases) that help to describe the main topics approached by the research. Our hypothesis here is that topics may play an important role in predicting the productivity of grants, since particular research lines might have higher levels of productivity~\citep{tohalino2020analyzing}.
    In order to analyze whether productivity has a dependency on the research subject, we considered two measurements taking into account the success history of different subjects. The \emph{global} importance considers the success of a subject in the whole dataset. Differently, the \emph{local} importance considers the success of subjects in grants conducted by the PI being analyzed. The success rate of a subject $X$ is computed as the number of productive grants approaching $X$  divided by the total number of grants associated with $X$. The criteria used to characterize productivity is detailed in Section \ref{methodology:dataset}. Three different sets of features were considered to represent the success of research subjects:

    \begin{enumerate}

        \item \emph{$subjFeat_A$}: each researcher was characterized using a vector summarizing the local and global success of the approached subjects. Because many subjects might be related to a PI in previous projects, we summarized the success rate of the approached subjects observed for keywords. In particular, we considered the average, the standard deviation and the maximum success rate observed for the keywords. Therefore, for each PI, six features were considered: both local and global strategies were used to compute the success rate, and three summarization strategies were applied.

        \item \emph{$subjFeat_B$}: we first obtained the $k$-most frequent subjects of the researcher. Then we calculated the global success rate and local success rate vectors for these subjects. The generated vectors were considered feature vectors. We evaluated with $k$ ranging between $10$ and $50$ subjects.

        \item \emph{$subjFeat_C$}: This version is similar to \emph{$subjFeat_A$}, but instead of considering the success rate, we considered the frequency count of the researcher's subjects.

    \end{enumerate}

\end{enumerate}

\subsection{Classification}\label{methodology:classification}

{
The main purpose of this study is to probe whether bibliometric information of researchers (e.g. their publication history and participation in previous research grants) are relevant factors to predict productivity of their research grants.
}
{In order to address the problem of class unbalancing in the classification scheme~\citep{li2010learning}, we considered a grant as productive if it yielded at least one publication. Thus, for each considered research field, the total number of positive (i.e. grants with at least one publication) and negative instances are more regularly distributed. This is compatible with previous related research~\citep{tohalino2020analyzing}.
If a higher threshold was considered to label an instance as positive, only a small percentage of grants would be considered as positive and this effect would lead to a high level of class unbalancing. The fraction of positive instances found for MED, DENT and VET are $41.6\%$, $49.5\%$ and $32.9\%$, respectively.}

To analyze the relationship between the extracted features and
grants productivity, we used the following machine learning algorithms: $k$-Neatest Neighbors (kNN), Support Vector Machines (SVM), Naive Bayes (NB), Neural Networks (MLP) and Decision Trees (DTrees). The algorithms hyperparameters were optimized using the strategy described in~\citep{amancio2014systematic,rodriguez2019clustering}.
The evaluation of the classification was based on the 10-fold cross-validation method to split the dataset into training and test datasets~\citep{duda2012pattern}. A  description of the used algorithms and the strategy used to balance the classes are described in Appendix \ref{sec:aa}.

\section{Results and discussion}\label{section:results}

In this section, we discuss the results we obtained when evaluating whether the considered features can be used to predict the productivity of grants.
We divided our analysis into the following sections: In Section~\ref{results:single}, we report the performance of different features when they are individually evaluated. In Section~\ref{results:combinations}, we discuss the results we obtained when researcher features are combined. The relevance of features for the classification task is also analyzed. In Section~\ref{results:voting}, we report the results obtained when combining several classifiers via voting method.

\subsection{Performance analysis of single features}\label{results:single}

\begin{figure}%
    \centering
    \includegraphics[width=11.5cm]{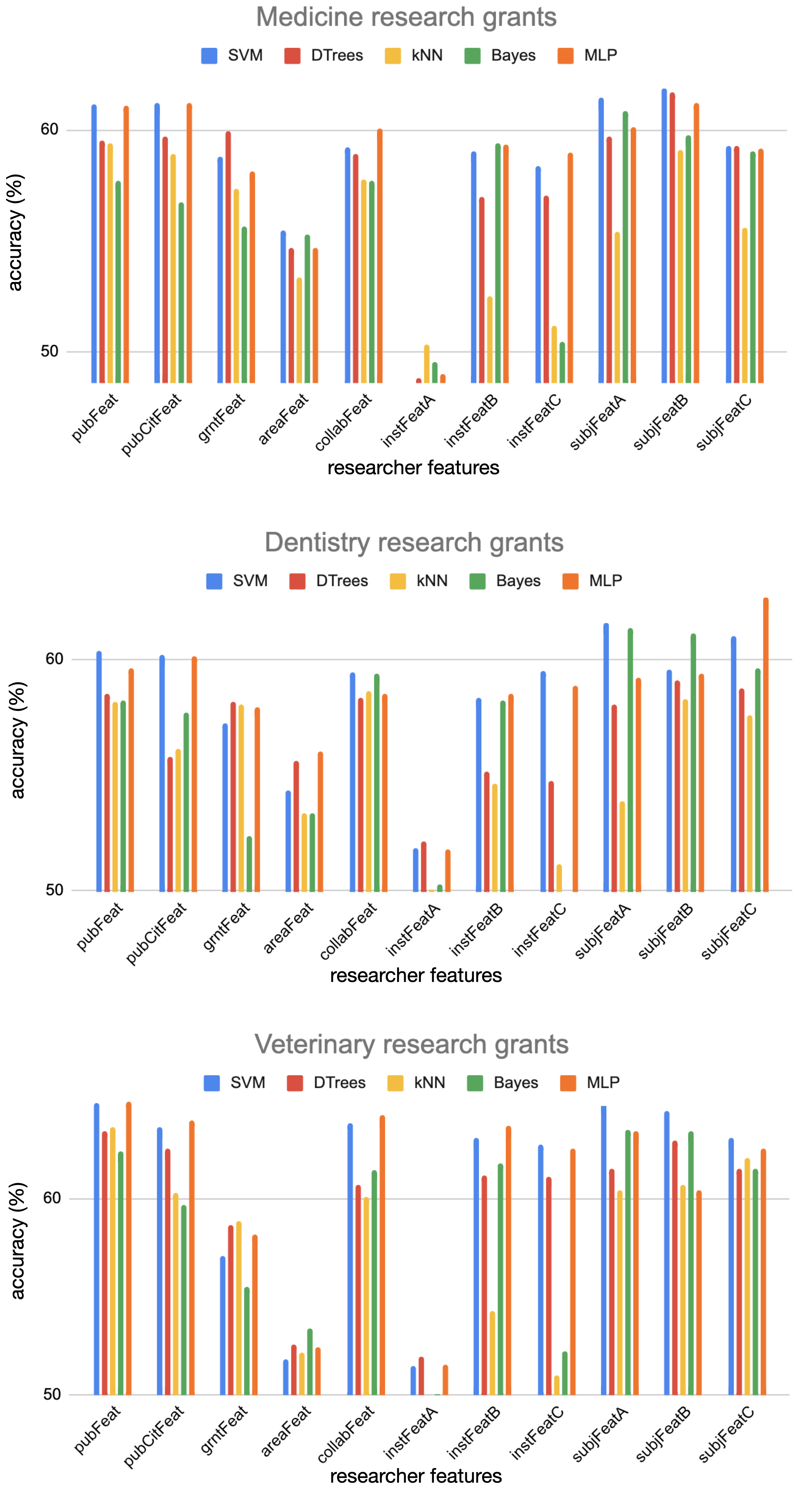}
    \caption{Results based on accuracy rate obtained from the evaluation of each researcher feature. We considered projects from the following areas: Medicine, Dentistry and Veterinary Medicine. From each researcher we considered seven features and their variations.}%
    \label{fig:results}%
\end{figure}

In this section, we discuss the results when each family of features is individually analyzed. The obtained results for each of the considered research areas are shown in Figure \ref{fig:results}. For the Medicine area, we observed that the best result was found with subject-based features ($subjFeat_B$), meaning that the average success of the topic being approached plays a role in predicting productivity. In the best scenario, the accuracy rate reached roughly $62\%$. While this is not a very high accuracy rate, this result turned out to be significant.
We also found that, for this research area, publication and citation history also play a role in predicting productivity. Both previous productivity (\emph{pubFeat}) and impact (\emph{pubCitFeat}) have a significant role in predicting grants productivity. All other features were found to be not statistically discriminative. Particularly, we found a very low discriminative performance for a particular institution feature ({\emph{instFeat}$_A$}) that considers the total number of projects hosted by the institution. Because this feature might be related to the size of the institution, this result suggests that being in a larger university is not necessarily linked to a higher productivity. Surprisingly we found that the diversity of areas can not be used as a source of productivity. While interdisciplinary researchers usually attain better performance in funding~\citep{sun2021interdisciplinary}, we did not observe any significant relationship between grants interdisciplinarity and productivity.

When analyzing the results obtained for the Dentistry area, we also found similar results. The productivity rate of the approached subject ($\emph{subjFeat}_C$) was found to be the most relevant feature to identify productive Dentistry grants. In this case, the highest accuracy rate reached $63.0\%$. We also found that publication and citation-based features also generated statistically significant results. All other features were found to be not significantly correlated with productivity.

The highest accuracy rates were found for the Veterinary area. An accuracy of roughly 66\% was obtained with the MLP and SVM methods. Once again we observe that the previous success of a topic is correlated with future success. In a similar fashion, publication and citation-based features also displayed significant accuracy rates. Differently from Medicine and Dentistry, a statistically significant relationship between collaboration features and productivity has also been found. This means that the number of previous PI's collaborators could be an indicator of
future grant productivity. A correlation was also found for two institution-based features ($\emph{instFeat}_B$ and $\emph{instFeat}_C$). While the total number of projects hosted by the PI's institution is not correlated with productivity, the fraction of productive grants has a higher correlation rate. Similarly, the total number of publications associated with the PI's institution also seems to be correlated with productivity.

Concerning the methods, the performance varied according to the considered feature and dataset. When considering only the best results across all variations of features and methods, we observed that the highest accuracy rate was found with SVM (Medicine and Veterinary Medicine) and MLP (Dentistry). The best results are highlighted in the Supplementary Information. However, the best results obtained for each method are very similar when considering a single feature classification. In general, the worst accuracy rates were obtained with kNN and Naive Bayes.

\subsection{Performance analysis of feature combinations}\label{results:combinations}

While in the previous section we analyzed the discriminability of each feature when they are used individually, here we analyze whether combinations of features can lead to optimized results. In addition to considering all features, we also considered other feature selection algorithms to find an optimized combination of attributes~\citep{kou2020evaluation}. The first approach used to combine features considered random subgroups of features. We considered feature sets of different sizes and we used them as input for the classification systems.
In Table \ref{tab:combinations}, we show the best accuracy rate found in each dataset, with different classifiers. We found that all best results were found to be statistically significant, and the best result being found for Veterinary Medicine (66.5\%). As observed in the single feature analysis, the results are not highly discriminative. In addition, the results show that the combination of features is not significantly better than the results found with single features, showing that no complementary information can be obtained with the selected machine learning methods. Regarding the methods, we note that SVM outperformed all other classifiers in all datasets. However, similar results were obtained with other classifiers, especially with Decision Trees.

\begin{table*}[h!]
\centering
\caption{Results based on accuracy rate obtained by performing feature combinations. We show the results from the best combination of features for each research area (Medicine, Dentistry and Veterinary Medicine).  The best results for each research area are highlighted. The SVM classifier always achieved the highest scores for all cases.}
\begin{tabular}{l|c|c|c}
\hline
\hline
\multirow{2}{*}{\bf Method} & {Medicine} & {Dentistry} & {Vet. Medicine} \\
    &  Accuracy (\%) &Accuracy (\%)  & Accuracy (\%) \\
\hline
DTrees   &  $61.96 \pm 1.12$ & $ 62.08 \pm 0.90$ & $65.46 \pm 1.59$ \\
SVM      &  ${\bf 62.82 \pm 0.67}$ & $ {\bf 62.50 \pm 1.22}$ & ${\bf 66.57 \pm 1.57}$ \\
kNN      &  $59.29 \pm 0.88$ & $ 61.58 \pm 0.91$ & $64.27 \pm 1.49$ \\
Bayes    &  $60.34 \pm 0.41$ & $ 61.00 \pm 0.37$ & $64.06 \pm 1.61$ \\
MLP    &  $60.99 \pm 0.61$ & $ 60.28 \pm 0.61$ & $64.22 \pm 0.82$ \\
\hline
\hline
\end{tabular}
\label{tab:combinations}
\end{table*}

In addition to the approach based on a random selection of features, we also used an approach based on the Gini coefficient~\citep{nembrini2018revival}. This approach is widely used to find relevant features in methods based on decision trees~\citep{pedregosa2011scikit} and has also been used in the scientometrics context~\citep{tohalino2020analyzing}. Using the Gini index, each feature was given a relevance value and then we selected the top $k$-features to analyze the gain in discriminability, with $k$ ranging between $10$ and $100$. Then, we used the selected features along with the SVM algorithm in order to perform the classification process. We considered the SVM classifier because it displayed the highest accuracy rates with the selected features.

The results obtained with the Gini feature selection are displayed in Table \ref{tab:relevance}. Overall the results show that there is no significant improvement in performance when one compares with the results obtained with the results displayed in Table \ref{tab:combinations}.  However, a small improvement can be observed mainly for the Veterinary Medicine Area. While this strategy was not useful to improve significantly the performance, this allowed an improved representation since similar results were obtained with a much smaller set of features (10 features in the case of Veterinary Medicine). We also note that that a large set of features -- even when selected via Gini method -- does note necessarily improve the discriminability rate.

\begin{table*}[h!]
\centering
\caption{Results obtained from the evaluation of feature selection methods. We tested with the most $10$, $20$, $50$ and $100$ relevant features. The best results for each research area are highlighted. The accuracy rate was obtained with the SVM method, since this method provided the best results.}
\begin{tabular}{l|c|c|c}
\hline
\hline
\multirow{2}{*}{\bf $k$ relevant features} & {Medicine} & {Dentistry} & {Vet. Medicine} \\
    &  Accuracy (\%) &Accuracy (\%)  & Accuracy (\%) \\
\hline
Top-10 features   &  $59.94 \pm 1.09$ & $  60.35 \pm 1.18$ & $67.35 \pm 1.28$ \\
Top-20 features    &  $60.24 \pm 1.00$ & $  60.68 \pm 0.52$ & ${\bf 67.37 \pm 1.14}$ \\
Top-50 features    &  $61.99 \pm 0.55$ & $  61.74 \pm 1.15$ & $66.98 \pm 1.55$ \\
Top-100 features    &  ${\bf 62.58 \pm 1.11}$ & $ {\bf 62.77 \pm 0.46}$ & $65.77 \pm 1.14$ \\
\hline
\hline
\end{tabular}
\label{tab:relevance}
\end{table*}

In addition to providing a compact representation, the feature selection algorithm allowed us to investigate which features are the most important for the classification task. This analysis is different from the analysis performed in the previous section because different discriminability performance can be observed when features are \emph{combined}~\citep{amancio2011comparing}. According to the Gini coefficient, the most relevant features for each field are:

\begin{itemize}

    \item \emph{Medicine}: institution-based and subject-based features displayed the highest Gini values. The success rate of the projects hosted by the PI's institution was a relevant feature in the family of features related to the institution ($\emph{instFeat}_C$). In a similar fashion, the success history associated with the approached subject was an important feature. This result suggests that, when used in combination with other features, the success history of both institution and approached subject are relevant to predict  the output of research projects.

    \item \emph{Dentistry}: $\emph{instFeat}_C$ and $\emph{subjFeat}_A$ were found to be the most relevant features. This result is similar to the one found for Medicine. The other variations related to subject-based features were also relevant: both history of global and local success of approached subjects were important for the task.

    \item \emph{Veterinary Medicine}: here the most relevant feature was the history of the PI's publications ($\emph{pubFeat}$). The other important variables were institution based features and
    all features based on the relevant of subject features ($\emph{subjFeat}_A$, $\emph{subjFeat}_B$ and $\emph{subjFeat}_C$).

\end{itemize}

The vast majority of relevant features are based on subject features (\emph{subjFeat}). Consequently, these results confirm the good performance obtained from the subject-based features when they are only considered for the classification systems (according to the results shown in Section~\ref{results:single}). We also observed that some variations of the features based on the institution of the researcher ($\emph{instFeat}$) have a degree of importance to characterize the performance of a researcher. However, it is important to recall that these features performed poorly when they were considered individually, while their performance improved when they were evaluated together with other features in a combined approach. These results indicate that institution-based could be an important factor, but individually they did not display a discriminative power. It is also interesting to note that the publication can also play a role in predicting success. This measurement turned out to be particularly important for the Veterinary Medicine area, even when used as a single feature (see Figure \ref{fig:results}).

\subsection{Performance analysis of ensembles: voting algorithm} \label{results:voting}

In the previous section, we analyzed if the combination of different features are able to improve the discriminability rates. Here we combine different classifiers to analyze if evidence from multiple methods can lead to optimized results. For this we used a voting algorithm~\citep{kiziloz2021classifier}. Two strategies were considered: (i) the use of all considered classifiers and (ii) the use of the best classifiers (see Material and Methods). According to the results presented in the previous sections, the classification systems that achieved the highest accuracy rates were the SVM and MLP algorithms, while Naive Bayes and kNN obtained the worst performance. We considered all the combinations of features described in the previous sections.

We show in Table~\ref{tab:voting} the results obtained from these evaluations.  The obtained results revealed that the combination of classifiers did not improve the results obtained in previous sections. In particular, using all classifiers is not useful given the low performance achieved especially by Naive Bayes and kNN. The performance of  SVM + MLP combination was also not useful to improve the performance of the classification.

\begin{table*}[h!]
\centering
\caption{Accuracy rates obtained from the evaluation of voting algorithms. For each research area we show the results of the following methods: All (when the results of all proposed classifiers are combined into the voting system) and Best (when only SVM and MLP are considered). The best results for each research area are highlighted.}
\begin{tabular}{l|c|c|c}
\hline \hline
\multirow{2}{*}{\bf Method} & {Medicine} & {Dentistry} & {Vet. Medicine} \\
    &  Accuracy (\%) &Accuracy (\%)  & Accuracy (\%) \\
\hline
All   &  $59.67 \pm 0.62$ & $ 57.88 \pm 0.36$ & $64.61 \pm 1.36$ \\
Best    &  ${\bf 62.07 \pm 0.89}$ & $ {\bf 62.50 \pm 0.74}$ & ${\bf 65.00 \pm 0.80}$ \\
\hline
\hline
\end{tabular}
\label{tab:voting}
\end{table*}

In sum, in the considered dataset, we found that combining different evidence from different features is more important than combining different classifiers. While some classifiers perform better than others, this result indicates that additional features could be used to improve the predictive power of the classifiers.
The study conducted here showed that the features we used are more relevant than features based on topical or complexity textual features~\citep{tohalino2020analyzing}. However, additional text information could also be obtained from  project abstracts and complement the characterization of research projects in order to improve the predictability of grants productivity. Additional text representations, including those based on network science~\citep{tohalino2018extractive} could also be used to characterize research grant texts.

\section{Conclusion} \label{sec:conc}

In this paper, we evaluated whether it is possible to predict the productivity of research projects by considering many different features to describe scientific entities.

Our analysis was conducted in three large subareas and considered grants awarded by \emph{São Paulo Research Foundation}, one of the largest research agencies in Brazil. We considered several features that could quantify the characteristics of research projects, PI's experience and the importance of host institutions. The relationship between the features and productivity was analyzed in the context of a traditional classification task.

Our analysis considered four different approaches to combine features and classifiers. First, we analyzed classifiers created with only a single feature. We then combined features via feature selection and relevance analysis. We also combined classifiers in a voting algorithm. Overall we found that the best results in all four considered approaches are statistically significant, meaning that some of the features play a role in predicting the output of research projects. The main results are summarized in Table \ref{tab:resumo}. All best results were found to be significant, though none of them reached a $70\%$ accuracy rate. The best results in different areas were obtained with distinct strategies. A single feature (the approached subject) was able to provide the highest accuracy rate for the Dentistry area.
The combination of features was able to provide the highest accuracy for Medicine. A feature selection algorithm finally provided the best results for Veterinary Medicine. Our analysis also revealed that the voting system combining evidence from multiple classifiers did not provide any improvement in classification performance.

\begin{table*}[h!]
\caption{Summary of the best results for each proposed approach. We highlighted the highest scores for each research area. The highest accuracy was found for the Veterinary Medicine field. Four different approaches were considered: (i) classification based on a single feature; (b) random combination of features; (iii) classification based on feature selection; and (iv) combination of classifiers via voting strategy.}
\begin{tabular}{l|c|c|c}
\hline
\hline
\multirow{2}{*}{\bf Method} & {Medicine} & {Dentistry} & {Vet. Medicine} \\
    &  Accuracy (\%) &Accuracy (\%)  & Accuracy (\%) \\
\hline
Single feature & $62.07 \pm 0.70$ & ${\bf 63.00 \pm 0.55}$ & $65.69 \pm 1.24$ \\
Features combination & ${\bf 62.82 \pm 0.67}$ & $62.50 \pm 1.22$ & $66.57 \pm 1.57$\\
Feature relevance & $62.58 \pm 1.11$ & $62.77 \pm 0.46$ & ${\bf 67.37 \pm 1.14}$  \\
Voting system &  $62.07 \pm 0.89$ & $62.50 \pm 0.74 $ & $ 65.00 \pm 0.80$ \\
\hline
\hline
\end{tabular}
\label{tab:resumo}
\end{table*}

While we found a dependency between the features and the output of projects, in all considered areas, the accuracy rates were not very high. This reinforces the fact that predicting the output of grants is not a trivial task and one can not rely only on machine learning to make predictions with high accuracy, at least with the considered features. The results found in this study were slightly better than the ones found using only textual features~\citep{tohalino2020analyzing}.

In future works, it would be interesting to analyze whether the use of wider contexts could lead to improved results. In text analysis, the access to the full content of research projects could provide more information than the title and abstract. A possible analysis could be the extraction of textual patterns via network analysis~\citep{marinho2016authorship,stella2021cognitive,amancio2012unveiling}. Unfortunately, full textual information is not currently available in our dataset.
Other extensions of this work could also be investigated. This includes other productivity and impact criteria, such as the total number of citations received by a grant, the reputation of journals and conferences associated with grant publications and other quality and impact criteria. We could also use collaborative network-based approaches~\citep{amancio2012use,correa2017patterns,amancio2015topological} to analyze whether scientific collaborations and team formation strategies may play a role in grants productivity.

\section*{Acknowledgments}
This study was financed in part by the Coordenação de Aperfeiçoamento de Pessoal de Nível Superior - Brasil (CAPES) - Finance Code 001.

\newpage

\bibliographystyle{abbrvnat}

\begin{thebibliography}{48}
\providecommand{\natexlab}[1]{#1}
\providecommand{\url}[1]{\texttt{#1}}
\expandafter\ifx\csname urlstyle\endcsname\relax
  \providecommand{\doi}[1]{doi: #1}\else
  \providecommand{\doi}{doi: \begingroup \urlstyle{rm}\Url}\fi

\bibitem[Acuna et~al.(2012)Acuna, Allesina, and Kording]{acuna2012predicting}
D.~E. Acuna, S.~Allesina, and K.~P. Kording.
\newblock Predicting scientific success.
\newblock \emph{Nature}, 489\penalty0 (7415):\penalty0 201--202, 2012.

\bibitem[Amancio et~al.(2011)Amancio, Altmann, Oliveira~Jr, and
  da~Fontoura~Costa]{amancio2011comparing}
D.~R. Amancio, E.~G. Altmann, O.~N. Oliveira~Jr, and L.~da~Fontoura~Costa.
\newblock Comparing intermittency and network measurements of words and their
  dependence on authorship.
\newblock \emph{New Journal of Physics}, 13\penalty0 (12):\penalty0 123024,
  2011.

\bibitem[Amancio et~al.(2012{\natexlab{a}})Amancio, Oliveira~Jr, and
  Costa]{amancio2012unveiling}
D.~R. Amancio, O.~N. Oliveira~Jr, and L.~d.~F. Costa.
\newblock Unveiling the relationship between complex networks metrics and word
  senses.
\newblock \emph{EPL (Europhysics Letters)}, 98\penalty0 (1):\penalty0 18002,
  2012{\natexlab{a}}.

\bibitem[Amancio et~al.(2012{\natexlab{b}})Amancio, Oliveira~Jr, and
  Costa]{amancio2012use}
D.~R. Amancio, O.~N. Oliveira~Jr, and L.~d.~F. Costa.
\newblock On the use of topological features and hierarchical characterization
  for disambiguating names in collaborative networks.
\newblock \emph{EPL (Europhysics Letters)}, 99\penalty0 (4):\penalty0 48002,
  2012{\natexlab{b}}.

\bibitem[Amancio et~al.(2014)Amancio, Comin, Casanova, Travieso, Bruno,
  Rodrigues, and da~Fontoura~Costa]{amancio2014systematic}
D.~R. Amancio, C.~H. Comin, D.~Casanova, G.~Travieso, O.~M. Bruno, F.~A.
  Rodrigues, and L.~da~Fontoura~Costa.
\newblock A systematic comparison of supervised classifiers.
\newblock \emph{PloS one}, 9\penalty0 (4):\penalty0 e94137, 2014.

\bibitem[Amancio et~al.(2015)Amancio, Oliveira~Jr, and
  Costa]{amancio2015topological}
D.~R. Amancio, O.~N. Oliveira~Jr, and L.~d.~F. Costa.
\newblock Topological-collaborative approach for disambiguating authors’
  names in collaborative networks.
\newblock \emph{Scientometrics}, 102\penalty0 (1):\penalty0 465--485, 2015.

\bibitem[Bagrow et~al.(2018)Bagrow, Berenberg, and Bongard]{bagrow2018neural}
J.~P. Bagrow, D.~Berenberg, and J.~Bongard.
\newblock Neural language representations predict outcomes of scientific
  research.
\newblock \emph{arXiv preprint arXiv:1805.06879}, 2018.

\bibitem[Bar-Ilan(2008)]{bar2008h}
J.~Bar-Ilan.
\newblock The h-index of h-index and of other informetric topics.
\newblock \emph{Scientometrics}, 75\penalty0 (3):\penalty0 591--605, 2008.

\bibitem[Bauder(2020)]{bauder2020international}
H.~Bauder.
\newblock International mobility and social capital in the academic field.
\newblock \emph{Minerva}, pages 1--21, 2020.

\bibitem[Boyack et~al.(2018)Boyack, Smith, and Klavans]{boyack2018toward}
K.~W. Boyack, C.~Smith, and R.~Klavans.
\newblock Toward predicting research proposal success.
\newblock \emph{Scientometrics}, 114\penalty0 (2):\penalty0 449--461, 2018.

\bibitem[Breiman(2001)]{breiman2001random}
L.~Breiman.
\newblock Random forests.
\newblock \emph{Machine learning}, 45\penalty0 (1):\penalty0 5--32, 2001.

\bibitem[Corr{\^e}a~Jr et~al.(2017)Corr{\^e}a~Jr, Silva, Costa, and
  Amancio]{correa2017patterns}
E.~A. Corr{\^e}a~Jr, F.~N. Silva, L.~d.~F. Costa, and D.~R. Amancio.
\newblock Patterns of authors contribution in scientific manuscripts.
\newblock \emph{Journal of Informetrics}, 11\penalty0 (2):\penalty0 498--510,
  2017.

\bibitem[Dietterich(2000)]{dietterich2000ensemble}
T.~G. Dietterich.
\newblock Ensemble methods in machine learning.
\newblock In \emph{International workshop on multiple classifier systems},
  pages 1--15. Springer, 2000.

\bibitem[Duda et~al.(2012)Duda, Hart, and Stork]{duda2012pattern}
R.~O. Duda, P.~E. Hart, and D.~G. Stork.
\newblock \emph{Pattern classification}.
\newblock John Wiley \& Sons, 2012.

\bibitem[Fortunato et~al.(2018)Fortunato, Bergstrom, B{\"o}rner, Evans,
  Helbing, Milojevi{\'c}, Petersen, Radicchi, Sinatra, Uzzi, Vespignani,
  Waltman, Wang, and Barab{\'a}si]{Fortunatoeaao0185}
S.~Fortunato, C.~T. Bergstrom, K.~B{\"o}rner, J.~A. Evans, D.~Helbing,
  S.~Milojevi{\'c}, A.~M. Petersen, F.~Radicchi, R.~Sinatra, B.~Uzzi,
  A.~Vespignani, L.~Waltman, D.~Wang, and A.-L. Barab{\'a}si.
\newblock Science of science.
\newblock \emph{Science}, 359\penalty0 (6379), 2018.
\newblock ISSN 0036-8075.
\newblock \doi{10.1126/science.aao0185}.
\newblock URL \url{https://science.sciencemag.org/content/359/6379/eaao0185}.

\bibitem[Franceschet and Costantini(2010)]{franceschet2010effect}
M.~Franceschet and A.~Costantini.
\newblock The effect of scholar collaboration on impact and quality of academic
  papers.
\newblock \emph{Journal of informetrics}, 4\penalty0 (4):\penalty0 540--553,
  2010.

\bibitem[Haykin(2008)]{2008haykinneural}
S.~O. Haykin.
\newblock \emph{Neural Networks and Learning Machines}.
\newblock 3rd Edition. Prentice Hall, 3 edition, 2008.
\newblock ISBN 0131471392,9780131471399.
\newblock URL
  \url{http://gen.lib.rus.ec/book/index.php?md5=0239F16656E6E5E7DB7AAA160CF9F854}.

\bibitem[He et~al.(2019)He, Zhen, and Wu]{he2019measuring}
Z.~He, N.~Zhen, and C.~Wu.
\newblock Measuring and exploring the geographic mobility of american
  professors from graduating institutions: Differences across disciplines,
  academic ranks, and genders.
\newblock \emph{Journal of Informetrics}, 13\penalty0 (3):\penalty0 771--784,
  2019.

\bibitem[Ioannidis et~al.(2020)Ioannidis, Boyack, and
  Baas]{ioannidis2020updated}
J.~P. Ioannidis, K.~W. Boyack, and J.~Baas.
\newblock Updated science-wide author databases of standardized citation
  indicators.
\newblock \emph{PLoS Biology}, 18\penalty0 (10):\penalty0 e3000918, 2020.

\bibitem[Kiziloz(2021)]{kiziloz2021classifier}
H.~E. Kiziloz.
\newblock Classifier ensemble methods in feature selection.
\newblock \emph{Neurocomputing}, 419:\penalty0 97--107, 2021.

\bibitem[Kong et~al.(2020)Kong, Zhang, Zhang, Bu, Ding, and Xia]{kong2020gene}
X.~Kong, J.~Zhang, D.~Zhang, Y.~Bu, Y.~Ding, and F.~Xia.
\newblock The gene of scientific success.
\newblock \emph{ACM Transactions on Knowledge Discovery from Data (TKDD)},
  14\penalty0 (4):\penalty0 1--19, 2020.

\bibitem[Kou et~al.(2020)Kou, Yang, Peng, Xiao, Chen, and
  Alsaadi]{kou2020evaluation}
G.~Kou, P.~Yang, Y.~Peng, F.~Xiao, Y.~Chen, and F.~E. Alsaadi.
\newblock Evaluation of feature selection methods for text classification with
  small datasets using multiple criteria decision-making methods.
\newblock \emph{Applied Soft Computing}, 86:\penalty0 105836, 2020.

\bibitem[Kumbure et~al.(2020)Kumbure, Luukka, and Collan]{kumbure2020new}
M.~M. Kumbure, P.~Luukka, and M.~Collan.
\newblock A new fuzzy k-nearest neighbor classifier based on the bonferroni
  mean.
\newblock \emph{Pattern Recognition Letters}, 140:\penalty0 172--178, 2020.

\bibitem[Larrimore et~al.(2011)Larrimore, Jiang, Larrimore, Markowitz, and
  Gorski]{larrimore2011peer}
L.~Larrimore, L.~Jiang, J.~Larrimore, D.~Markowitz, and S.~Gorski.
\newblock Peer to peer lending: The relationship between language features,
  trustworthiness, and persuasion success.
\newblock \emph{Journal of Applied Communication Research}, 39\penalty0
  (1):\penalty0 19--37, 2011.

\bibitem[Letchford et~al.(2016)Letchford, Preis, and
  Moat]{letchford2016advantage}
A.~Letchford, T.~Preis, and H.~S. Moat.
\newblock The advantage of simple paper abstracts.
\newblock \emph{Journal of Informetrics}, 10\penalty0 (1):\penalty0 1--8, 2016.

\bibitem[Li et~al.(2010)Li, Liu, and Hu]{li2010learning}
D.-C. Li, C.-W. Liu, and S.~C. Hu.
\newblock A learning method for the class imbalance problem with medical data
  sets.
\newblock \emph{Computers in biology and medicine}, 40\penalty0 (5):\penalty0
  509--518, 2010.

\bibitem[Lu et~al.(2019)Lu, Bu, Dong, Wang, Ding, Larivi{\`e}re, Sugimoto,
  Paul, and Zhang]{lu2019analyzing}
C.~Lu, Y.~Bu, X.~Dong, J.~Wang, Y.~Ding, V.~Larivi{\`e}re, C.~R. Sugimoto,
  L.~Paul, and C.~Zhang.
\newblock Analyzing linguistic complexity and scientific impact.
\newblock \emph{Journal of Informetrics}, 13\penalty0 (3):\penalty0 817--829,
  2019.

\bibitem[Marinho et~al.(2016)Marinho, Hirst, and
  Amancio]{marinho2016authorship}
V.~Q. Marinho, G.~Hirst, and D.~R. Amancio.
\newblock Authorship attribution via network motifs identification.
\newblock In \emph{2016 5th Brazilian Conference on Intelligent Systems
  (BRACIS)}, pages 355--360. IEEE, 2016.

\bibitem[Markowitz(2019)]{markowitz2019words}
D.~M. Markowitz.
\newblock What words are worth: National science foundation grant abstracts
  indicate award funding.
\newblock \emph{Journal of Language and Social Psychology}, 38\penalty0
  (3):\penalty0 264--282, 2019.

\bibitem[McCallum et~al.(1998)McCallum, Nigam, et~al.]{mccallum1998comparison}
A.~McCallum, K.~Nigam, et~al.
\newblock A comparison of event models for naive bayes text classification.
\newblock In \emph{AAAI-98 workshop on learning for text categorization},
  volume 752, pages 41--48. Citeseer, 1998.

\bibitem[Nembrini et~al.(2018)Nembrini, K{\"o}nig, and
  Wright]{nembrini2018revival}
S.~Nembrini, I.~R. K{\"o}nig, and M.~N. Wright.
\newblock The revival of the gini importance?
\newblock \emph{Bioinformatics}, 34\penalty0 (21):\penalty0 3711--3718, 2018.

\bibitem[Note1()]{Note1}
Note1.
\newblock https://fapesp.br/en/about.

\bibitem[Note2()]{Note2}
Note2.
\newblock https://bv.fapesp.br/pt/.

\bibitem[Paiva et~al.(2012)Paiva, Lima, and Paiva]{paiva2012articles}
C.~E. Paiva, J.~P. d. S.~N. Lima, and B.~S.~R. Paiva.
\newblock Articles with short titles describing the results are cited more
  often.
\newblock \emph{Clinics}, 67\penalty0 (5):\penalty0 509--513, 2012.

\bibitem[Pedregosa et~al.(2011)Pedregosa, Varoquaux, Gramfort, Michel, Thirion,
  Grisel, Blondel, Prettenhofer, Weiss, Dubourg, et~al.]{pedregosa2011scikit}
F.~Pedregosa, G.~Varoquaux, A.~Gramfort, V.~Michel, B.~Thirion, O.~Grisel,
  M.~Blondel, P.~Prettenhofer, R.~Weiss, V.~Dubourg, et~al.
\newblock Scikit-learn: Machine learning in python.
\newblock \emph{the Journal of machine Learning research}, 12:\penalty0
  2825--2830, 2011.

\bibitem[Rafols and Meyer(2010)]{rafols2010diversity}
I.~Rafols and M.~Meyer.
\newblock Diversity and network coherence as indicators of interdisciplinarity:
  case studies in bionanoscience.
\newblock \emph{Scientometrics}, 82\penalty0 (2):\penalty0 263--287, 2010.

\bibitem[Rinia et~al.(2002)Rinia, van Leeuwen, and van Raan]{rinia2002impact}
E.~Rinia, T.~van Leeuwen, and A.~van Raan.
\newblock Impact measures of interdisciplinary research in physics.
\newblock \emph{Scientometrics}, 53\penalty0 (2):\penalty0 241--248, 2002.

\bibitem[Rodriguez et~al.(2019)Rodriguez, Comin, Casanova, Bruno, Amancio,
  Costa, and Rodrigues]{rodriguez2019clustering}
M.~Z. Rodriguez, C.~H. Comin, D.~Casanova, O.~M. Bruno, D.~R. Amancio, L.~d.~F.
  Costa, and F.~A. Rodrigues.
\newblock Clustering algorithms: A comparative approach.
\newblock \emph{PloS one}, 14\penalty0 (1):\penalty0 e0210236, 2019.

\bibitem[Ruta and Gabrys(2005)]{ruta2005classifier}
D.~Ruta and B.~Gabrys.
\newblock Classifier selection for majority voting.
\newblock \emph{Information fusion}, 6\penalty0 (1):\penalty0 63--81, 2005.

\bibitem[Salatino et~al.(2017)Salatino, Osborne, and Motta]{salatino2017topics}
A.~A. Salatino, F.~Osborne, and E.~Motta.
\newblock How are topics born? understanding the research dynamics preceding
  the emergence of new areas.
\newblock \emph{PeerJ Computer Science}, 3:\penalty0 e119, 2017.

\bibitem[Silva et~al.(2016)Silva, Amancio, Bardosova, Costa, and
  Oliveira~Jr]{silva2016using}
F.~N. Silva, D.~R. Amancio, M.~Bardosova, L.~d.~F. Costa, and O.~N.
  Oliveira~Jr.
\newblock Using network science and text analytics to produce surveys in a
  scientific topic.
\newblock \emph{Journal of Informetrics}, 10\penalty0 (2):\penalty0 487--502,
  2016.

\bibitem[Siudem et~al.(2020)Siudem, {\.Z}oga{\l}a-Siudem, Cena, and
  Gagolewski]{siudem2020three}
G.~Siudem, B.~{\.Z}oga{\l}a-Siudem, A.~Cena, and M.~Gagolewski.
\newblock Three dimensions of scientific impact.
\newblock \emph{Proceedings of the National Academy of Sciences}, 2020.

\bibitem[Stella et~al.(2021)Stella, Swanson, Hills, and
  Teixeira]{stella2021cognitive}
M.~Stella, T.~Swanson, T.~T. Hills, and A.~S. Teixeira.
\newblock Cognitive network science as a framework for detecting structural
  patterns and emotions in suicide letters.
\newblock 2021.

\bibitem[Sun et~al.(2021)Sun, Livan, Ma, and Latora]{sun2021interdisciplinary}
Y.~Sun, G.~Livan, A.~Ma, and V.~Latora.
\newblock Interdisciplinary researchers attain better performance in funding.
\newblock \emph{arXiv preprint arXiv:2104.13091}, 2021.

\bibitem[Tohalino et~al.(2021)Tohalino, Quispe, and
  Amancio]{tohalino2020analyzing}
J.~A.~V. Tohalino, L.~V.~C. Quispe, and D.~R. Amancio.
\newblock Analyzing the relationship between text features and research
  proposal productivity.
\newblock \emph{Scientometrics}, 126:\penalty0 4255–4275, 2021.
\newblock \doi{10.1007/s11192-021-03926-x}.

\bibitem[Tohalino and Amancio(2018)]{tohalino2018extractive}
J.~V. Tohalino and D.~R. Amancio.
\newblock Extractive multi-document summarization using multilayer networks.
\newblock \emph{Physica A: Statistical Mechanics and its Applications},
  503:\penalty0 526--539, 2018.

\bibitem[Wang et~al.(2019)Wang, Jones, and Wang]{wang2019early}
Y.~Wang, B.~F. Jones, and D.~Wang.
\newblock Early-career setback and future career impact.
\newblock \emph{Nature communications}, 10\penalty0 (1):\penalty0 1--10, 2019.

\bibitem[Zeng et~al.(2017)Zeng, Shen, Zhou, Wu, Fan, Wang, and
  Stanley]{zeng2017science}
A.~Zeng, Z.~Shen, J.~Zhou, J.~Wu, Y.~Fan, Y.~Wang, and H.~E. Stanley.
\newblock The science of science: From the perspective of complex systems.
\newblock \emph{Physics Reports}, 714:\penalty0 1--73, 2017.

\end{thebibliography}

\newpage

\appendix

\section{List of Supervised Classifiers} \label{sec:aa}

In this section we describe the main classifiers used to predict the productivity of research grants. In addition to the traditional machine learning algorithms, we also used a technique to combine all considered pattern recognition methods.

\begin{enumerate}

    \item \emph{$k$-Neatest Neighbors} (kNN): With the aim of classifying an unknown element from dataset, the kNN method first selects the $k$-nearest elements from the training dataset. Then, the category assigned to the unknown element corresponds to the majority class which is detected in the selected $k$-set. $k$ is an important parameter of the algorithm and is chosen via optimization methods~\citep{amancio2014systematic}.

    \item \emph{Support Vector Machines} (SVM): Given the training examples, this method constructs a hyperplane with the aim of finding a separation between the classes of the dataset. This method has several parameters, including parameters that sets the kernel function used to create a hyperplane~\citep{amancio2014systematic}.

    \item \emph{Naive Bayes} (NB): This classifier is a supervised learning algorithm the uses Bayes' theorem with a strong assumption that features are independent~\citep{mccallum1998comparison}. In this sense, the following equation is used to predict the class $\hat{y}$:
    \begin{equation}
        \label{eq:naive}
        \hat{y} = \underset{y}{\operatorname{argmax}}~ P(y)\prod\limits_{i=1}^n P(x_i|y)
    \end{equation}
    where $x_i$ is a feature. The training dataset can be used to compute the probabilities $P(y)$ and $P(x_i|y)$. Additional parameters related to this method and the optimization process are described elsewhere~\citep{amancio2014systematic}.

    \item \emph{Multilayer Perceptron} (MLP): This method is based on a neural network model considering one or more hidden layers which has a training process that usually involves the Back-Propagation algorithm~\citep{2008haykinneural}. Two main hyper-parameters exists in this model: (i) the number of layers; and (ii) the number of neurons for each layer. These parameters can also be chosen via optimization.

    \item \emph{Decision trees}: decision tree methods create models that able to predict the value of a target variable by learning decision rules. The rules
    are inferred from several input variables, i.e. the features. A typical decision tree comprises nodes and edges, where nodes represent features and edges represent the decision chosen for each attribute. Each internal with children is labeled with some input feature, while  leaf nodes are labeled with a class. The classification process starts at the root node and ends when a leaf node is reached. As the decision walks through nodes, a rule is applied and the decision on that rule guides the choice of the children to be chosen as next step. Once the decision reaches a leaf node, the predicted label corresponds to the respective label stored in that node~\citep{breiman2001random}

    The choice of features that are evaluated in each node depends on the feature relevance metric that decides which feature best discriminates the dataset. One important relevance metric is the Gini impurity~\citep{nembrini2018revival}, a metric that has already been used to evaluate the relevance of features in the context of productivity prediction~\citep{tohalino2020analyzing}.
    The Gini impurity measures how often a randomly selected instance from the dataset would be mislabeled if it was randomly classified according to the distribution of the categories in the considered subset.
    A feature is considered significant in a given node if the test performed using that feature results in a decrease in the Gini impurity. The relevance of a feature is obtained by averaging  the decrease in impurity computed in all tree nodes using the considered feature.

    \item \emph{Ensemble learning}: ensemble methods combine the predictions of several machine learning algorithms in order to obtain a better predictive performance over a single method~\citep{dietterich2000ensemble}. Most ensemble methods build several estimators independently, and then they average the predictions of each method. Usually the voting method is a simple, yet effective approach designed to combine the predictions from several supervised classifiers. In this approach, the input classifiers are trained and tested independently. Then the observed predictions from all classifiers are combined by using a majority vote to predict the class labels. Therefore, the class receiving the highest number of votes is chosen as the final predicted class~\citep{ruta2005classifier}.
    When a draw occurs, we used the membership strength~\citep{kumbure2020new} provided by each classifier to make a decision.

\end{enumerate}

\section{Results obtained from the evaluation of each research feature}

\begin{table*}[h!]
\centering
\caption{Accuracy rate obtained when considering the classification with single features. We considered projects from the following areas: Medicine, Dentistry and Veterinary Medicine. The best results for each classifier are highlighted.}
\begin{tabular}{l|c|c|c|c|c}
\hline
\hline
\multicolumn{1}{c|}{\multirow{3}{*}{\bf Features}} & \multicolumn{5}{c}{Research Projects on \emph{Medicine}} \\
    &  \multicolumn{1}{c}{DTrees} & \multicolumn{1}{c}{SVM}  & \multicolumn{1}{c}{kNN} & \multicolumn{1}{c}{Bayes} & \multicolumn{1}{c}{MLP} \\
        &  \multicolumn{1}{c}{Accuracy (\%) } & \multicolumn{1}{c}{Accuracy (\%)}  & \multicolumn{1}{c}{Accuracy (\%)} & \multicolumn{1}{c}{Accuracy (\%)} & \multicolumn{1}{c}{Accuracy (\%)} \\
\hline
$\emph{pubFeat}$	    & $59.48 \pm 0.64$ & $61.24 \pm 0.40$ & ${\bf 59.36 \pm 0.81}$ & $57.51 \pm 0.43$ & $61.21 \pm 0.53 $ \\
$\emph{pubCitFeat}$	    & $59.67 \pm 0.85$ & $61.36 \pm 0.36$ & $58.80 \pm 0.35$ & $56.56 \pm 0.42$ & ${\bf 61.37 \pm 0.39} $ \\
$\emph{grntFeat}$	    & $59.91 \pm 0.75$ & $58.68 \pm 0.63$ & $57.14 \pm 0.92$ & $55.43 \pm 0.53$ & $57.98 \pm 1.31 $ \\
$\emph{areaFeat}$	    & $54.45 \pm 0.86$ & $55.25 \pm 0.93$ & $53.13 \pm 1.11$ & $55.05 \pm 0.39$ & $54.46 \pm 0.95 $ \\
$\emph{collabFeat}$	    & $58.84 \pm 0.93$ & $59.14 \pm 1.06$ & $57.63 \pm 0.73$ & $57.51 \pm 0.42$ & $60.06 \pm 0.99 $ \\
$\emph{instFeat}_A$  & $48.94 \pm 1.25$ & $48.73 \pm 1.48$ & $50.28 \pm 0.05$ & $49.59 \pm 0.47$ & $49.09 \pm 1.16 $ \\
$\emph{instFeat}_B$    & $56.80 \pm 0.62$ & $58.94 \pm 0.67$ & $52.33 \pm 2.24$ & $59.37 \pm 0.81$ & $59.30 \pm 0.74 $ \\
$\emph{instFeat}_C$    & $56.85 \pm 1.13$ & $58.25 \pm 0.46$ & $51.08 \pm 0.53$ & $50.40 \pm 0.43$ & $58.90 \pm 0.49 $ \\
$\emph{subjFeat}_A$	    & $59.70 \pm 0.53$ & $61.61 \pm 0.34$ & $55.17 \pm 2.12$ & ${\bf 60.95 \pm 0.44}$ & $60.16 \pm 0.94 $ \\
$\emph{subjFeat}_B$	    & ${\bf 61.89 \pm 0.93}$ & ${\bf 62.07 \pm 0.70}$ & $59.05 \pm 0.83$ & $59.77 \pm 0.36$ & $61.32 \pm 1.02 $ \\
$\emph{subjFeat}_C$	    & $59.24 \pm 0.92$ & $59.24 \pm 0.76$ & $55.35 \pm 1.98$ & $58.93 \pm 0.60$ & $59.06 \pm 0.65 $ \\
\hline
\multicolumn{1}{c|}{\multirow{2}{*}{\bf Features}} & \multicolumn{5}{c}{Research Projects on \emph{Dentistry}} \\
    &  \multicolumn{1}{c}{DTrees} & \multicolumn{1}{c}{SVM}  & \multicolumn{1}{c}{kNN} & \multicolumn{1}{c}{Bayes} & \multicolumn{1}{c}{MLP} \\
\hline
$\emph{pubFeat}$	    & $58.40 \pm 0.96$ & $60.41 \pm 0.26$ & $58.05 \pm 1.03$ & $58.09 \pm 0.40$ & $59.56 \pm 0.68 $ \\
$\emph{pubCitFeat}$	    & $55.57 \pm 0.86$ & $60.24 \pm 0.11$ & $55.90 \pm 0.39$ & $57.52 \pm 0.17$ & $60.12 \pm 0.56 $ \\
$\emph{grntFeat}$	    & $58.02 \pm 0.59$ & $57.03 \pm 0.54$ & $57.92 \pm 0.50$ & $52.18 \pm 0.55$ & $57.78 \pm 0.45 $ \\
$\emph{areaFeat}$	    & $55.35 \pm 0.65$ & $54.11 \pm 1.01$ & $53.12 \pm 0.68$ & $53.11 \pm 0.14$ & $55.81 \pm 0.83 $ \\
$\emph{collabFeat}$	    & $58.20 \pm 0.87$ & $59.38 \pm 0.45$ & ${\bf 58.50 \pm 0.98}$ & $59.32 \pm 0.33$ & $58.39 \pm 0.61 $ \\
$\emph{instFeat}_A$	    & $51.97 \pm 0.40$ & $51.67 \pm 0.42$ & $50.04 \pm 0.14$ & $50.21 \pm 0.15$ & $51.61 \pm 0.41 $ \\
$\emph{instFeat}_B$	    & $54.92 \pm 0.57$ & $58.20 \pm 0.56$ & $54.37 \pm 1.43$ & $58.09 \pm 0.39$ & $58.36 \pm 0.23 $ \\
$\emph{instFeat}_C$	    & $54.48 \pm 0.87$ & $59.46 \pm 0.56$ & $51.05 \pm 0.30$ & $49.93 \pm 0.31$ & $58.74 \pm 0.67 $ \\
$\emph{subjFeat}_A$	    & $57.90 \pm 1.10$ & ${\bf 61.74 \pm 0.48}$ & $53.63 \pm 1.01$ & ${\bf 61.53 \pm 0.17}$ & $59.12 \pm 0.43 $ \\
$\emph{subjFeat}_B$	    & ${\bf 59.01 \pm 0.61}$ & $59.50 \pm 0.65$ & $58.13 \pm 0.83$ & $61.22 \pm 0.23$ & $59.34 \pm 1.04 $ \\
$\emph{subjFeat}_C$	    & $58.61 \pm 1.16$ & $61.13 \pm 0.74$ & $57.41 \pm 0.78$ & $59.59 \pm 0.33$ & ${\bf 63.00 \pm 0.55}$ \\
\hline
\multicolumn{1}{c|}{\multirow{2}{*}{\bf Features}} & \multicolumn{5}{c}{Research Projects on \emph{Veterinary Medicine}} \\
    &  \multicolumn{1}{c}{DTrees} & \multicolumn{1}{c}{SVM}  & \multicolumn{1}{c}{kNN} & \multicolumn{1}{c}{Bayes} & \multicolumn{1}{c}{MLP} \\
\hline
$\emph{pubFeat}$	    & ${\bf 63.82 \pm 1.45}$ & $65.58 \pm 0.76$ & ${\bf 64.10 \pm 1.27}$ & $62.68 \pm 0.62$ & ${\bf 65.67 \pm 0.72} $ \\
$\emph{pubCitFeat}$	    & $62.82 \pm 0.95$ & $64.08 \pm 0.88$ & $60.33 \pm 0.95$ & $59.66 \pm 0.71$ & $64.49 \pm 1.14 $ \\
$\emph{grntFeat}$	    & $58.50 \pm 1.81$ & $56.90 \pm 1.58$ & $58.75 \pm 2.40$ & $55.30 \pm 0.69$ & $58.00 \pm 1.46 $ \\
$\emph{areaFeat}$	    & $52.37 \pm 1.95$ & $51.66 \pm 2.27$ & $52.01 \pm 1.66$ & $53.19 \pm 1.04$ & $52.27 \pm 2.46 $ \\
$\emph{collabFeat}$	    & $60.75 \pm 1.53$ & $64.31 \pm 2.00$ & $60.11 \pm 0.96$ & $61.61 \pm 1.34$ & $64.85 \pm 1.51 $ \\
$\emph{instFeat}_A$	    & $51.80 \pm 1.10$ & $51.34 \pm 0.96$ & $49.98 \pm 0.39$ & $50.04 \pm 0.45$ & $51.42 \pm 1.17 $ \\
$\emph{instFeat}_B$	    & $61.26 \pm 1.07$ & $63.45 \pm 1.26$ & $54.02 \pm 3.25$ & $61.95 \pm 0.93$ & $64.14 \pm 1.04 $ \\
$\emph{instFeat}_C$	    & $61.18 \pm 1.16$ & $63.07 \pm 1.43$ & $50.89 \pm 0.45$ & $52.07 \pm 2.94$ & $62.87 \pm 0.81 $ \\
$\emph{subjFeat}_A$	    & $61.63 \pm 1.44$ & ${\bf 65.69 \pm 1.24}$ & $60.44 \pm 2.39$ & ${\bf 63.96 \pm 1.09}$ & $63.88 \pm 1.88 $ \\
$\emph{subjFeat}_B$	    & $63.33 \pm 1.63$ & $65.10 \pm 1.48$ & $60.78 \pm 2.38$ & $63.82 \pm 1.42$ & $60.45 \pm 1.24 $ \\
$\emph{subjFeat}_C$	    & $61.69 \pm 0.85$ & $63.49 \pm 1.48$ & $62.30 \pm 0.93$ & $61.66 \pm 1.07$ & $62.83 \pm 1.96 $ \\
\hline
\hline
\end{tabular}
\label{tab:single}
\end{table*}

\end{document}